\documentclass[aps,pre,twocolumn,groupedaddress,showpacs,amssymb,amsmath]{revtex4-1}

\usepackage{graphicx}
\usepackage{tabularx}
\usepackage{color}
\usepackage{amsmath}
\usepackage{comment}
\newcommand{\be}{\begin{equation}}
\newcommand{\ee}{\end{equation}}
\newcommand{\bea}{\begin{eqnarray}}
\newcommand{\eea}{\end{eqnarray}}
\newcommand{\ep}{\epsilon}
\usepackage{dcolumn}
\usepackage{hyperref}
\usepackage{bm}
\usepackage{epsf}
\usepackage{subfigure}
\usepackage{braket}
\usepackage{amsfonts}

\def\ep{{\epsilon}}
\def\la{{\langle}}
\def\ra{{\rangle}}
\def\lad{{\langle \langle}}
\def\rad{{\rangle \rangle}}
\def\om{{\omega}}

\begin{document}
\title{ Universal bounds on cooling power and cooling efficiency for autonomous absorption refrigerators} 
\author{Sandipan Mohanta}
\affiliation{Department of Physics, Indian Institute of Science Education and Research Pune, Dr. Homi Bhabha Road, Ward No. 8, NCL Colony, Pashan, Pune, Maharashtra 411008, India}
\author{Sushant Saryal}
\affiliation{Department of Physics, Indian Institute of Science Education and Research Pune, Dr. Homi Bhabha Road, Ward No. 8, NCL Colony, Pashan, Pune, Maharashtra 411008, India}
\author{Bijay Kumar Agarwalla}
\email{bijay@iiserpune.ac.in}
\affiliation{Department of Physics, Indian Institute of Science Education and Research Pune, Dr. Homi Bhabha Road, Ward No. 8, NCL Colony, Pashan, Pune, Maharashtra 411008, India}

\date{\today}
\begin{abstract}


For steady-state autonomous absorption refrigerators operating in the linear response regime, we show that there exists a hierarchy between the relative fluctuation of currents for cold, hot, and work terminals. Our proof requires the Onsager's reciprocity relation along with the refrigeration condition that sets the direction of the mean currents for each terminal. As a consequence, the universal bounds on the mean cooling power, obtained following the thermodynamic uncertainty relations, receive a hierarchy. Interestingly, within this hierarchy, the tightest bound is given in terms of the work current fluctuation.  Furthermore,  the relative uncertainty hierarchy hands over additional bounds that can be tighter than the bounds obtained from the thermodynamic uncertainty relations.
Interestingly, all of these bounds saturate in the tight-coupling limit. We test the validity of our results for two paradigmatic absorption refrigerator models: (i) a four-level working fluid and (ii) a two-level working fluid, operating, respectively, in the weak (additive) and in the strong (multiplicative) system-bath interaction regime. 


\end{abstract}
\maketitle

{\it Introduction.--}  Autonomous absorption refrigerator (AR) is a three terminal setup that operates in non-equilibrium steady state and continuously direct energy to flow from cold $(c)$ to hot  $(h)$ terminal by absorbing energy from the ultra-hot work $(w)$ terminal  \cite{QAR-review-kos, Mitchison,QAR-early-1,QAR-early-2,QAR-early-3,QAR-Correa-1,QAR-Correa-2,QAR-Correa-3,QAR-agar1,Segal-abs-1,Segal-abs-2,Segal-abs-3,Segal-abs-4,Jordan,QAR-Brask,chen,Scarani,Popescu,Alonso}.  The first useful AR for industrial application was realised dates back in the 19th century by the Carr\'e brothers \cite{Gordon, first}.
With the rapid advancement in quantum technologies, intense efforts are now directed towards understanding and realising smallest possible ARs \cite{QAR-early-1, QAR-early-2, QAR-early-3,QAR-Correa-1,QAR-Correa-2, QAR-agar1,QAR-Correa-3,chen,QAR-Brask,Popescu} that can operate with maximum possible cooling efficiency and power, by taking advantage of possible quantum resources.
Various proposals to realise quantum ARs using platforms such as superconducting qubits, arrays of quantum dots have been put forward \cite{ QAR-expt-p1,QAR-expt-p2,QAR-expt-p3,QAR-expt-PRL},
with one successful experimental implementation using trapped ions is achieved very recently \cite{QAR-expt1}.


Notably, as the system size shrinks, the fluctuations (quantum and/or thermal) can play an important role in deciding the ultimate fate on the performance of such thermal machines. For the past two decades, the field of stochastic and quantum thermodynamics \cite{st-thermo1,st-thermo2,st-thermo3,Q-thermo2, Q-thermo3, fluc-1,fluc-2,fluc-3,sto-Benenti,esp-Q-thermo} are continuously providing a fundamental and deep understanding about such questions \cite{Keiji-engine, trade-off-engine,Benenti-fluc}.  In this regard, {\it thermodynamic uncertainty relations} (TURs) \cite{Barato:2015:UncRel,trade-off-engine,Gingrich:2016:TUP,Falasco,Garrahan18,Timpanaro,TUR-Goold,Saito-TUR,Junjie-TUR,supriya,Agarwalla-TUR,TUR-open,Agarwalla-TUR-2}, a trade-off bound between relative fluctuation of currents (precision) and the entropy production rate (cost), have provided universal upper bounds on the performance of thermal machines in terms of the current fluctuations and are in fact tighter than the standard Carnot bound \cite{trade-off-engine}. More specifically, for multi-affinity driven thermal engines, it was shown that the TURs for different currents (heat, work) can lead to different bounds on the output power \cite{trade-off-engine}. However, it was not apparent whether or not there exists any relationship between these bounds. 


In this Letter, we show that for time-reversal symmetric ARs, relative fluctuation of currents for different terminals are not independent but follow a strict hierarchy in the linear response regime. As a result, the different universal bounds for the cooling power that are obtained following the TURs also receive a strict hierarchy. The tightest bound within this hierarchy is given when the bound is expressed in terms of the work current fluctuation. We further report additional bounds for both the cooling power and the cooling efficiency which can become tighter than the bounds that follow from the TURs.

The plan of this Letter is as follows: We first introduce a generic AR setup and provide a proof for the hierarchy in relative fluctuation for currents following the Onsager's reciprocity relation and further imposing conditions on the mean currents to realise an AR. We then discuss the bounds on efficiency and power that results from this hierarchy and further connect our results with the bounds that follow from the TURs. We provide numerical results and discuss these bounds for two paradigmatic quantum AR models with working medium consisting of a (i) four-level system and a (ii) two-level system, operating respectively in the weak and the strong system-bath coupling regime. 


{\it Absorption refrigerator and bounds .--} A generic setup for an autonomous absorption refrigerator (AR) consists of a working fluid (classical or quantum) connected to three thermal baths, namely the cold $(c)$, hot $(h)$  and work $(w)$ terminals. The inverse temperatures ($\beta_{\alpha} = 1/k_B T_{\alpha}$) of these terminals follow a sequence  $\beta_c > \beta_h >\beta_w$.  Given this setting, the primary task of an AR is to direct energy to flow from the cold  to the hot terminal via absorbing energy from the hottest work terminal and as a result, cooling down the cold terminal. The central objective of our work is to provide universal bounds on the cooling power $\la j_c \ra$ ($\la j_{\alpha}\ra$ is the mean current for terminal $\alpha$) and the cooling efficiency, defined as, $\la  \ep \ra = \langle j_c \rangle / \la j_w \ra$. Throughout the paper, we follow the convention that currents flowing into the working fluid from different terminals are positive. 

Before presenting our main results, we first highlight the bounds on cooling efficiency and power that can be obtained following the universal TURs. To receive these bounds we follow Ref.~(\cite{trade-off-engine}). For an out-of-equilibrium system in steady-state, subjected to multiple thermodynamic affinities, the TURs \cite{Barato:2015:UncRel,trade-off-engine,Gingrich:2016:TUP,Falasco,Garrahan18,Timpanaro,TUR-Goold,Saito-TUR,Junjie-TUR,supriya,Agarwalla-TUR,TUR-open,Agarwalla-TUR-2} provide  lower bounds on the relative fluctuation of individual currents (heat, work, particle etc.) in terms of the net entropy production rate.  These bounds therefore dismiss the possibility of optimizing the relative fluctuations and the dissipation in an arbitrary manner. For the AR setup, the TUR relations can be written down for individual currents as \cite{Barato:2015:UncRel}
\be
\la \sigma \ra \frac{D_{\alpha}}{\la j_{\alpha} \ra^2} \geq 1, \quad \alpha=c, h, w
\label{TUR-bounds}
\ee
where $\la \sigma \ra=- \sum_{\alpha} \beta_{\alpha} \la j_{\alpha} \ra$ is the net entropy production (EP) rate in the steady-state. For the multi-affinity AR setup, the equality here corresponds to the tight-coupling limit \cite{supp}. In the above expression, the dispersion $D_{\alpha}$ is related to the second cumulant of current by $ D_{\alpha} = \lad j_{\alpha}^2 \rad/2$. Note that the above bound is universal in the linear response regime and very recently its validity was questioned beyond this regime and a looser bound was reported \cite{Junjie-TUR, TUR-Goold}. In this work, we focus our attention only in the linear response regime. Writing down the TUR bound for cold terminal
and using the expression for EP rate, we immediately obtain an universal bound on the cooling efficiency in terms of the mean current and fluctuation of the cold terminal as \cite{supp}
\bea
\langle \epsilon \rangle &\leq& \epsilon_{\rm cool} \Big{/} \Big(1 + \frac{\la j_c \ra }{(\beta_c\!-\!\beta_h) \, D_c}\Big) \equiv {\cal E}_c,
\label{eff-bound-udo}
\eea
which is tighter than the standard three-terminal version of maximum cooling efficiency $\epsilon_{\rm cool}=\frac{(\beta_h-\beta_w)}{(\beta_c-\beta_h)}$ \cite{supp}. Note that, $\epsilon_{\rm cool}$
reduces to the standard two-terminal cooling efficiency version $(1 - \eta_c)/\eta_c$ in the limit ${\beta_w \ll \beta_h}$ with $\eta_c = \beta_h/(\beta_c-\beta_h)$.  
Similarly, using Eq.~(\ref{TUR-bounds}) for all the three terminals, three separate upper bounds on the cooling power can be received as
\bea
  && \la j_c \ra \leq (\beta_c \!-\! \beta_h) \frac{\Big(\ep_{\rm cool} \!-\! \la \ep \ra\Big)}{\la \ep \ra} \, D_c \equiv  {\cal P}_c, \nonumber \\
  && \la j_c \ra \leq (\beta_c \!-\! \beta_h) \Big(\ep_{\rm cool} \!-\! \la \ep \ra\Big) \frac{\la \ep \ra}{\Big (1 + \la \ep \ra\Big)^2} \, D_h  \equiv  {\cal P}_h, \nonumber \\
 && \la j_c \ra \leq  \big(\beta_c \!-\! \beta_h\big) \, \Big(\ep_{\rm cool} \!-\! \la \ep \ra\Big ) \la \ep \ra \,  D_w \equiv {\cal P}_w. 
  \label{power-bound-udo}
  \eea
Note that, the TUR does not provide any apparent relationship between these bounds for cooling power. 
In what follows, we show that, interestingly, these bounds follow a strict hierarchy that emerges once we impose the refrigeration condition by setting the direction of the mean currents for each terminal.  Generally speaking, our universal result follows from the hierarchy of the bounds that exists between the relative fluctuation of currents involving all the three terminals,
\be
\frac{D_c}{\la j_c \ra^2} \geq \frac{D_h}{\la j_h \ra^2} \geq \frac{D_w}{\la j_w \ra^2}.
\label{hei}
\ee
To prove this central result, we follow the arguments presented in Ref.~\cite{Universal-Sushant}. Let us first focus on the cold and the hot terminals. In the linear response limit, we can express the currents in terms of the Onsager's response coefficients as
\bea
\la j_{c} \ra &=& L_{cc} \,A^{w}_c+ L_{ch}\, A_h^{w}, \nonumber \\
\la j_h \ra &=& L_{hc} \, A^{w}_c+ L_{hh} \, A_h^{w}, \nonumber 
\label{onsager-ch}
\eea
where $A^{w}_{\alpha}= (\beta_w\!-\!\beta_{\alpha}),\, \alpha=c, h$ is the thermodynamic affinity, defined with respect to the temperature of the work terminal $(w)$ which is set as the reference in this particular case. Since we focus on time-reversal symmetric engines,  the off-diagonal Onsager coefficients satisfy $L_{ch} = L_{hc}$. Recall that, as per our convention, current flowing towards the working fluid is considered as positive. To prove the hierarchy, given in Eq.~(\ref{hei}), we construct the ratio for the relative fluctuations for cold and hot currents and define ${\cal Q}_{\rm ch} = \big[\frac{D_c}{\la j_c \ra^2}\big]/ \big[\frac{D_h}{\la j_h \ra^2}\big] $  which in the linear response regime can be expressed as,
\bea
\!{\cal Q}_{\rm ch} \!=\!  1\!+\! \frac{1}{L_{hh} \la j_c \rangle^2} \Big[L_{cc} L_{hh} \!\!-\!\! L_{hc}^2\Big] \Big[L_{hh} {\big(A_{h}^{w}\big)}^2\!\!-\! \!L_{cc} {\big(A_{c}^{w}\big)}^2\Big], \nonumber \\
\label{ch-1}
\eea
where we approximate the fluctuation in the linear response limit as  $D_{\alpha} = L_{\alpha \alpha}/2$ . Notice that, the positivity of the EP rate i.e., $\la \sigma \ra =\la j_c \ra A^{w}_{c}  + \la j_h \ra A^{w}_{h} \geq 0$ ensures that the determinant of the Onsager's matrix i.e., $ \Big[L_{cc} L_{hh} - L_{hc}^2\Big] \geq 0$. 
Now, interestingly, it turns out that under the operative condition as a refrigerator, the other term in Eq.~(\ref{ch-1}) is also always non-negative. In the refrigerator regime, one demands $\langle j_c \rangle >0 $ and $\la j_h \ra <0$ which implies $\langle j_c \rangle A_c^w<0$  and $\langle j_h \rangle A_h^w >0$ as the temperatures of the terminals follow the sequence  $\beta_c > \beta_h >\beta_w$. These two conditions together in the linear response regime provide the required inequality $\Big[L_{hh} {\big(A_{h}^{w}\big)}^2\!-\!L_{cc} {\big(A_{c}^{w}\big)}^2\Big]> 0$.  We therefore receive 
${\cal Q}_{ch} \geq 1$, i.e.,
\be
\frac{D_c}{\la j_c \ra^2} \geq \frac{D_h}{\la j_h \ra^2}.
\label{bound-ch}
\ee
The equality in the above bound corresponds to vanishing determinant of the Onsager's matrix which is achieved in the tight-coupling limit (all mean currents are proportional to each other). It is important to note that, although the condition on the mean work current $\la j_w \ra$ does not appear explicitly in the above proof, the requirement that current flowing from cold to hot terminal i.e.,  $\la j_c \ra >0$ and $\la j_h \ra <0$ immediately implies that energy is absorbed from the work terminal, i.e., $\la j_w \ra >0$, which would otherwise violate the second-law of thermodynamics. 
Following exactly the above steps, one can construct ${\cal Q}_{\rm h w}= \big[\frac{D_h}{\la j_c \ra^2}\big]/ \big[\frac{D_w}{\la j_w \ra^2}\big] $ involving the relative fluctuations for hot and work currents and receive,
\bea
\!{\cal Q}_{\rm hw} \!=\!  1\!+\! \frac{1}{\tilde{L}_{ww} \la j_h \rangle^2} \Big[\tilde{L}_{ww} \tilde{L}_{hh} \!\!-\!\! \tilde{L}_{hw}^2\Big] \Big[ \tilde{L}_{ww} {\big(A_{w}^{c}\big)}^2\!\!-\! \!\tilde{L}_{hh} {\big(A_{h}^{c}\big)}^2\Big], \nonumber \\
\eea
where the thermodynamic affinity $A^{c}_{\alpha}= (\beta_c\!-\!\beta_{\alpha}), \, \alpha= h, w$ is now defined keeping the cold bath temperature as the reference. The symbol tilde on the response coefficients indicate that these are now computed with respect to the affinities $A^c_{\alpha}$. Once again, if we fix the directions for the hot and work currents i.e., $\la j_h \ra <0$ and $\la j_w \ra >0$,
we receive the inequality $\Big[\tilde{L}_{ww} {\big(A_{w}^{c}\big)}^2\!\!-\! \!\tilde{L}_{hh} {\big(A_{h}^{c}\big)}^2\Big]>0$, and therefore, 
\be
\frac{D_h}{\la j_h \ra^2} \geq \frac{D_w}{\la j_w \ra^2}.
\label{bound-hw}
\ee
Note that, the above inequality is valid independent of the direction of current from the cold terminal, as in this case, imposing the conditions $\la j_h \ra <0$ and $\la j_w \ra >0$ does not fix a definite direction for $\la j_c \ra$. Therefore, focusing only in the refrigerator regime, Eq.~(\ref{bound-ch}) and Eq.~(\ref{bound-hw}) provides the required hierarchy, as given in Eq.~(\ref{hei}). This is the first central result of this paper. We would like to stress here that the central result presented in Eq.~(\ref{hei}) is completely independent from the TUR results. The crucial difference in our case arises by imposing additional constrains on the direction of the currents to realise a refrigerator. 
We now list down a few immediate consequences of the above result: 

(i) The inequality involving cold and  work terminals provides a completely different universal lower bounds for the cooling efficiency and cooling power 
\bea
\langle \ep \rangle &\equiv& \frac{\la j_c\ra}{\la j_w \ra} \leq \sqrt{\frac{D_c}{D_w}} \leq \ep_{\rm cool}, \nonumber \\
\langle j_c \rangle &\leq & \sqrt{\frac{D_c}{D_w}} \la j_w \ra  \leq \ep_{\rm cool} \, \la j_w \ra,
\label{bound-bj}
 \eea
compared to the set of universal bounds obtained in Eq.~(\ref{eff-bound-udo}) and Eq.~(\ref{power-bound-udo}) via TUR. It can be shown that, this bound is also lower than the maximum cooling efficiency $\epsilon_{\rm cool}$ \cite{supp}.

(ii) It is interesting to note that, because of the hierarchy in relative fluctuations, given in Eq.~(\ref{hei}),  three separate and independent looking bounds for cooling power, received from TUR (Eq.~(\ref{power-bound-udo})), are not really independent but follow a similar strict hierarchy. Since, 
 \be
 D_w \leq \frac{D_h}{\big(1+ \la \ep \ra\big)^2} \leq  \frac{D_c}{\la \ep \ra^2},
 \ee
we receive, 
\bea
\langle j_c \rangle &\leq &  {\cal P}_w \leq {\cal P}_h \leq {\cal P}_c, 
\eea
i.e., within this hierarchy, the cooling power gets most tightly bounded when the bound is expressed in terms of the fluctuations in the work terminal $D_w$. 
Interestingly, the additional bound for cooling power in Eq.~(\ref{bound-bj}) does not fit within this hierarchy. More importantly, the additional universal bounds 
in Eq.~(\ref{bound-bj}) can become tighter compared to bounds that follow from the TUR in Eq.~(\ref{eff-bound-udo}) and Eq.~(\ref{power-bound-udo}) and are therefore important to consider. 
Interestingly, as the equality for the TURs and the hierarchy of relative fluctuations is achieved in the tight-coupling (TC) limit, all the above bounds then saturate and we can write 
\bea
\langle \ep \rangle&=&  \sqrt{\frac{D_c}{D_w}}= {\cal E}_c \leq \ep_{\rm cool} \nonumber \\
\langle j_c \rangle &=&  \sqrt{\frac{D_c}{D_w}}  \la j_w \ra =  {\cal P}_w = {\cal P}_h = {\cal P}_c.
\label{bound-tight}
\eea
This is the second central result of this paper. In what follows, we test the validity of these bounds for two paradigmatic absorption refrigerator setups. 

%
%
\begin{figure}
\includegraphics[trim=10 120 10 120, clip, width=\columnwidth]{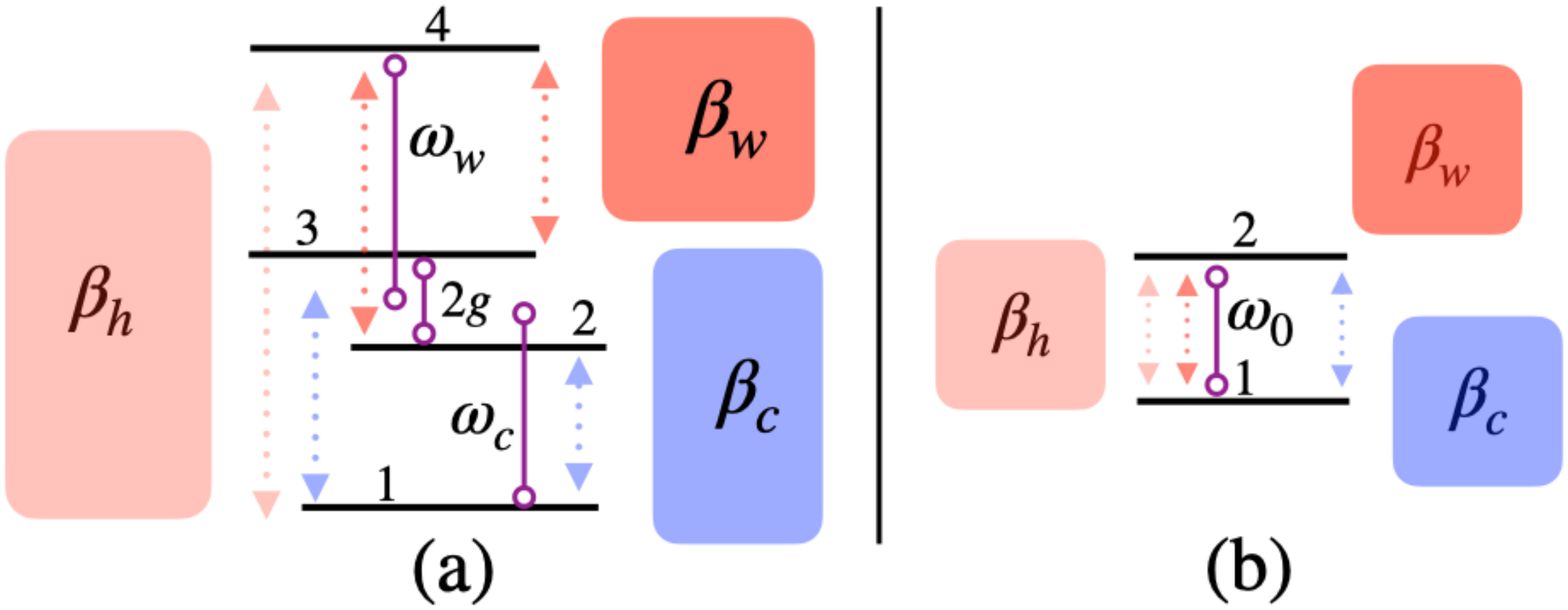}
\caption{(Color online):  (a) Model 1: Schematic of a four-level AR coupled weakly and (b) Model 2: Schematic of a two-level AR coupled strongly with all the three terminals. The inverse temperature of the baths follow the sequence $\beta_c > \beta_h > \beta_w$. The baths induced incoherent transitions are shown by the dashed arrows. The energy gaps between the states are indicated by solid lines. In the population dynamics, the different baths induced transition rates contributes in the additive manner for model 1 in (a) and in a multiplicative manner for model 2  in (b).}
\label{fig-QAR}
\end{figure}

{\it Example 1.-- Four-level system as the working fluid with weak (additive) system-bath coupling:}  As a first model example, we consider a four-level quantum AR setup which is weakly connected to three baths ($c, h, w$).  Bath induced incoherent transitions between the system states are displayed in Fig.~(\ref{fig-QAR}(a)). Note that, because of the weak system-bath coupling, these transition rates appear in an additive manner in the population dynamics. Interestingly, this particular working fluid goes beyond the traditional three level setup, the minimal model to realise a Quantum AR in the weak (tight)-coupling limit \cite{QAR-review-kos, QAR-Correa-1,QAR-Correa-2,supp}. 
The four-level model was recently studied by some of us by performing full-counting analysis for currents for the cold and the work terminals \cite{Segal-abs-4, Universal-Sushant}. The analysis can be easily extended to study statistics involving all three terminals. In the weak-coupling regime, following the standard quantum maser equation (QME) approach, a first-order differential equation for the characteristic function (CF) $ |{\cal Z}(\chi, t)\rangle$, corresponding to the counting-fields dressed population dynamics can be obtained as,
%
\bea \frac{ d| {\cal Z}(\chi, t)\rangle}{dt}=
\hat {\cal W}(\chi)|{\cal Z}(\chi, t)\rangle,
\label{eq:ZW}
\eea
where counting parameters $\chi=(\chi_w,\chi_h, \chi_c)$ keep track of the energy flowing out from the respective terminals. Here $\hat{\cal W}(\chi)$ is the  dressed rate matrix, given as,
%
%
\begin{widetext}
\begin{equation}
\hat {\cal W}(\chi)=
\begin{pmatrix}
-k_{1\to 2}^{c} -k_{1\to 3}^{c} -k_{1\to 4}^h & k_{2\to 1}^{c}e^{-i\chi_c(\omega_c-g)} & k_{3\to 1}^{c} e^{-i\chi_c(\omega_c+g)}& k_{4\to 1}^h e^{-i\chi_h(\omega_c+\omega_w)}\\
k_{1\to 2}^{c}e^{i\chi_c(\om_c-g)} &-k_{2\to 1}^{c} -k_{2\to 4}^w&0 & k_{4\to 2}^we^{-i\chi_w(\om_w+g)} \\
k_{1\to 3}^{c} e^{i\chi_c(\om_c+g)} &0& -k_{3\to 1}^{c} -k_{3\to 4}^w & k_{4\to 3}^w e^{-i\chi_w(\om_w-g)}\\
k_{1\to 4}^h e^{i\chi_h(\om_c+\om_w)} &  k_{2\to 4}^w e^{i\chi_w(\om_w+g)}   &  k_{3\to 4}^w e^{i\chi_w(\om_w-g)}& -k_{4\to 1}^h - k_{4\to 2}^w -k_{4\to 3}^w 
\end{pmatrix}
\end{equation}
\end{widetext}
with $k_{i\to j}^{\alpha}$ being the transition rate between states $i$ and $j$ due to terminal $\alpha$. The energy levels are labelled from bottom to top as 1 to 4 with energies given by $\epsilon_{1,2,3,4}=(0, \omega_c-g, \omega_c+g, \omega_c+\omega_w)$. For $\chi=0$, the rate matrix describes the standard population dynamics for the system states. Note that, for simplicity, in this example, we ignore the possible presence of quantum coherence in the system.
This however does not affect the main message presented in this work. 
The steady-state cumulant generating function (CGF) hands over the cumulants for currents which can be received from the CF as
\bea
{\cal G}(\chi) =  \lim_{t \to \infty} \ \frac{1}{t}\ln \langle I| {\cal Z}(\chi,t) \rangle
\eea
with $\langle I | = \langle 1 1 1 1| $ being a unit vector for the 4-level QAR.  The mean currents and the fluctuations are determined by taking the first and second derivative of the CGF with respect to the respective counting fields.  
\begin{figure}
\includegraphics[trim=45 0 45 0, clip, width=\columnwidth]{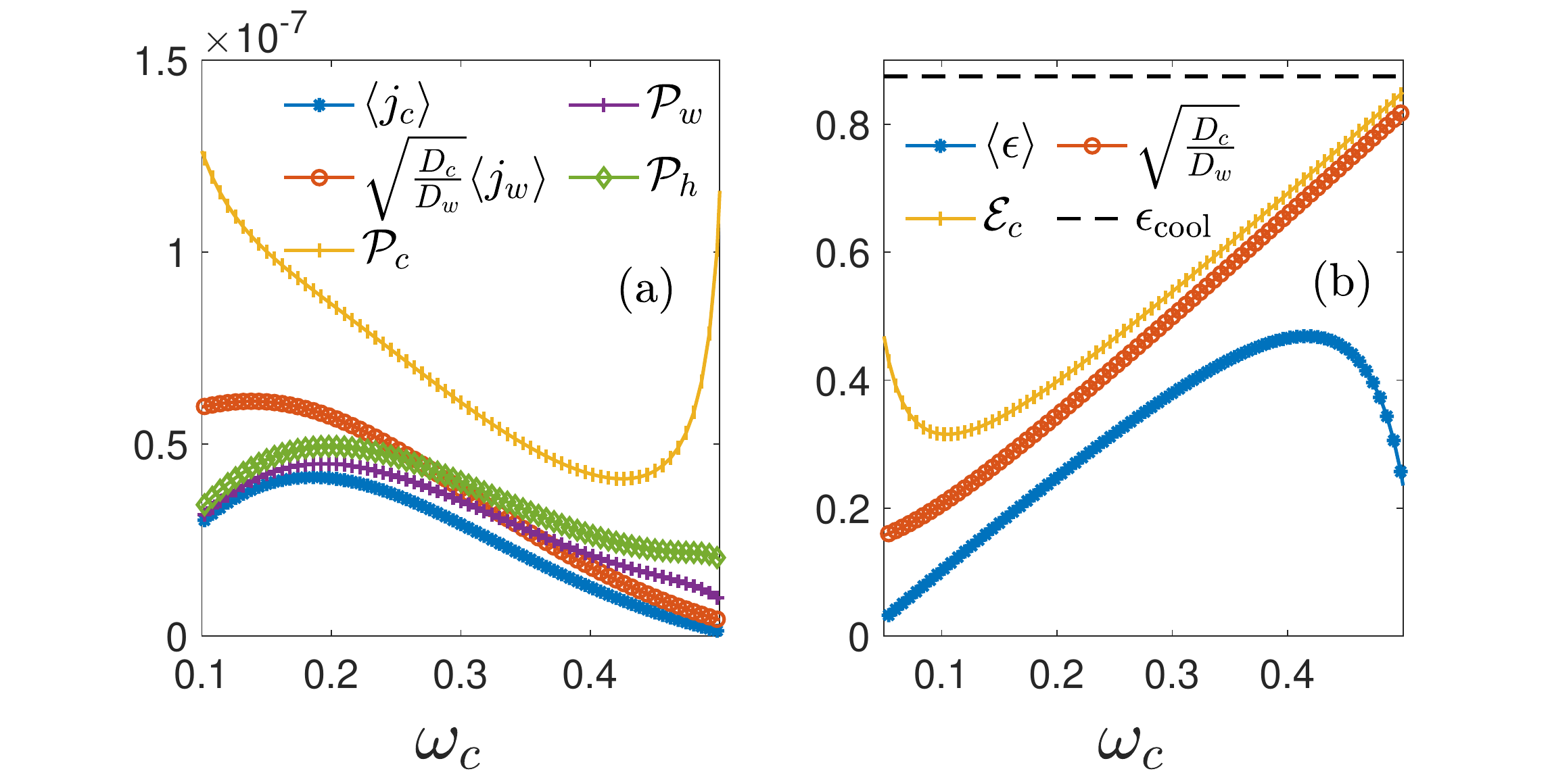}
\includegraphics[trim=40 0 40 0, clip, width=\columnwidth]{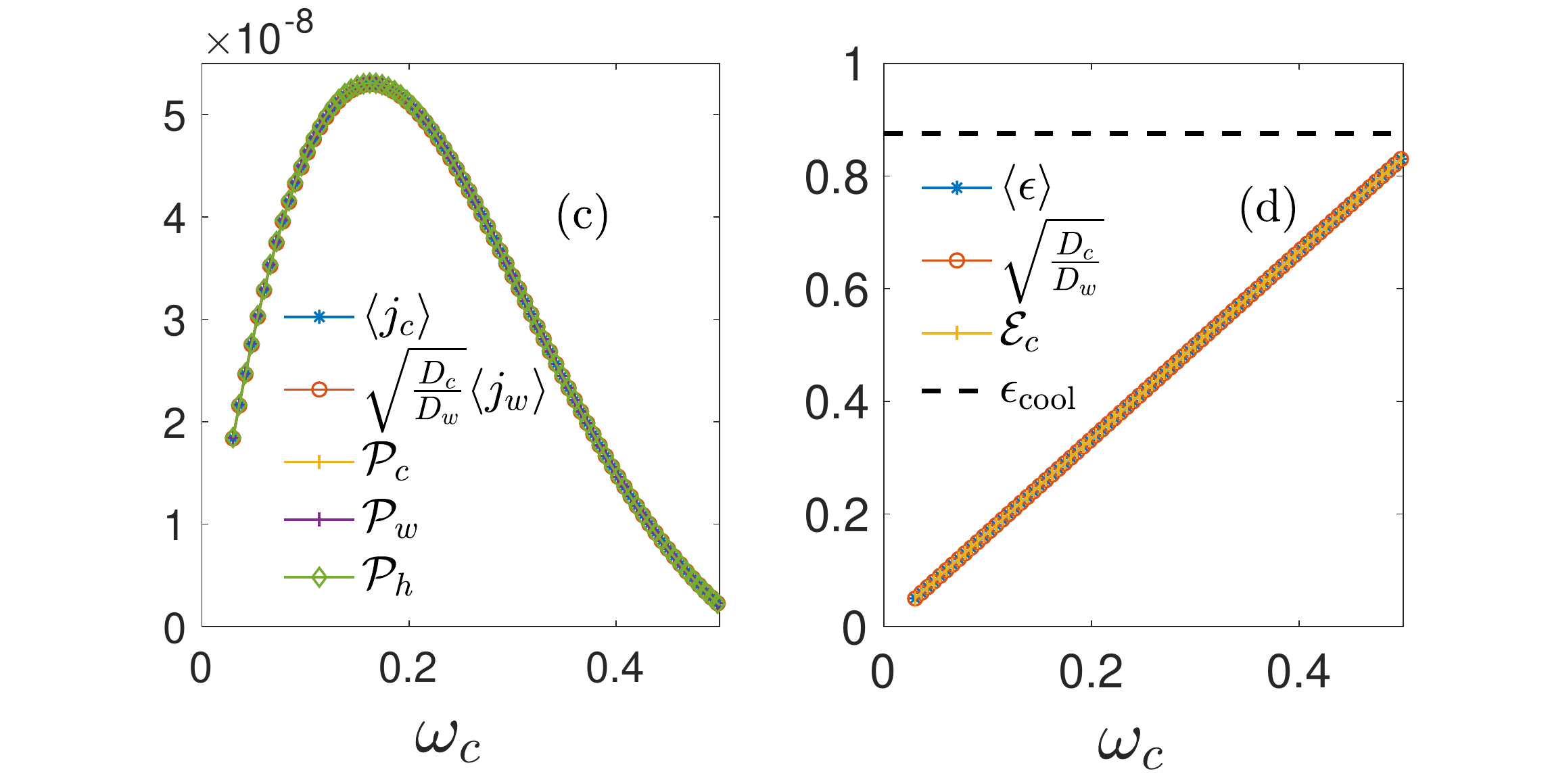}
\caption{(Color online): {Comparison between different bounds both for cooling power $\la j_c \ra$ ((a) and (c))  and cooling efficiency  $\la \ep \ra$  ((b) and (d)) for the four-level quantum AR as a function of $\omega_c$. The parameters chosen here are $T_c=0.14, T_h=0.15, T_w=0.16, \omega_w=0.6, \gamma_c=\gamma_h=\gamma_w=10^{-3}, \Lambda =50$. For (a) and (b)  $g=5\times 10^{-2}$ and for (c) and (d)  $g=1\times 10^{-4}$, corresponding to the tight-coupling situation. For these parameters, $\epsilon_{\rm cool} =0.8750$.}}
\label{four-level-QAR}
\end{figure}
In Fig.~(\ref{four-level-QAR}) we display the different bounds for cooling power and cooling efficiency. For the simulation, we choose the rate constants as 
$k_{i\to j}^{\alpha} = \Gamma_{\alpha}(\epsilon_j\!-\!\epsilon_i)\, n_{\alpha}(\epsilon_j\!-\!\epsilon_i)$  
with $n_{\alpha}(\Delta)=\left(e^{\beta_{\alpha}\Delta}-1\right)^{-1}$ as the Bose-Einstein distribution function at inverse temperature $\beta_{\alpha}$ and $\Gamma_{\alpha}(\Delta)=\gamma_{\alpha}\, \Delta \,e^{-|\Delta|/\Lambda}$ is the spectral density function of the bath $\alpha$ \cite{Universal-Sushant} with $\Lambda$ being the cutoff frequency and is assumed here to be large $\Lambda \gg \omega_{\alpha}, 1/\beta_{\alpha}$. 
Fig.~(\ref{four-level-QAR})((a) and (b)) shows results beyond the tight-coupling regime which corresponds to value of $g$ comparable to $\omega_{\alpha}$. 
The hierarchy in the bounds for cooling power ( ${\cal P}_w > {\cal P}_h > {\cal P}_c$ ) is clearly observed with the tightest bound given by ${\cal P}_w$. Note that, as mentioned earlier, the additional bound given in Eq.~(\ref{bound-bj}) do not follow the above hierarchy and can become a tighter bound, as is clearly seen in Fig.~(\ref{four-level-QAR})(a) for $\omega_c> 0.4$. Interestingly, in this parameter regime, the additional bound for cooling efficiency is also found to be  tighter than the bound predicted from the TUR in Eq.~(\ref{eff-bound-udo}).
Fig.~(\ref{four-level-QAR}) ((c) and (d)) shows results in the tight-coupling limit corresponding to closing of the gap between the states 2 and 3 ($g=1 \times 10^{-4} \ll \omega_{\alpha}$). In this case, as predicted in Eq.~(\ref{bound-tight}), all the bounds for efficiency and power saturate and coalesce with their respective exact mean values.




{\it Example 2.-- Two-level (qubit) system as the working fluid under strong (multiplicative) system-bath coupling:} The universal bounds obtained in our study are not limited to the nature of the system-bath coupling. To see this, as a second example, we choose another central model for  quantum AR where the working fluid consists of a two-level system but now coupled strongly to all three terminals (see Fig.~(\ref{fig-QAR}(b)). Interestingly, a qubit coupled weakly to all three terminals fails to work as a  quantum AR \cite{QAR-agar1} but supports the same functionality in the strong coupling regime, thanks to the non-additive nature of the transition rates in the rate equations, leading to a collective and simultaneous effect involving all the three terminals.
The full Hamiltonian mimicking a strong system-bath interaction can be written down as \cite{QAR-agar1},
\begin{figure}
\includegraphics[trim=15 255 15 255, clip, width=\columnwidth]{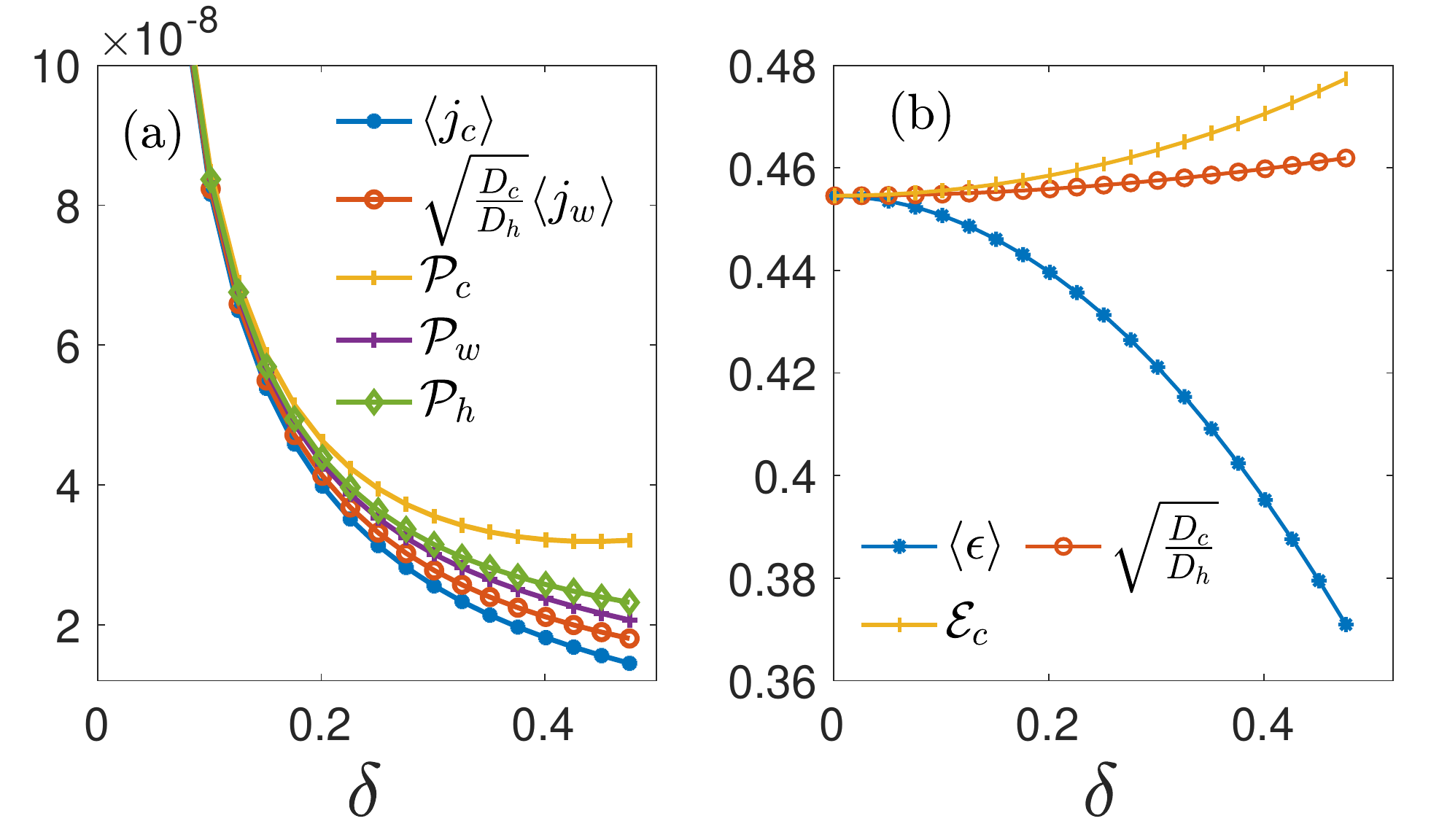}
\caption{(Color online): Comparison between different bounds for (a)  cooling power $\la j_c \ra$ and (b) cooling efficiency  $\la \ep \ra$ for two-level quantum AR operating under strong system-bath coupling as a function of $\delta$, as defined in Eq.~(\ref{spectra-2level}) . The parameters chosen here are $\omega_0=0$, $T_c=1, T_h=1.3, T_w=1.6, \omega_c=0.5, \omega_w = 1.1, \omega_h =\omega_c + \omega_w = 1.6, \gamma_h=\gamma_w=10^{-2}, \gamma_c=1$. For these set of parameters, $\epsilon_{\rm cool}=0.6250$. }
\label{2-level-QAR}
\end{figure}
\be
H= \frac{\omega_0}{2} \sigma_z + \frac{\sigma_x}{2} \otimes \big( B_c \otimes  B_h \otimes  B_w\big) + \sum_{\alpha=c, h,w} H_{B, \alpha}
\label{H-strong}
\ee
where the crucial second term involves the bath operators in a multiplicative manner and is responsible for inducing a collective effect. In other words, this particular term is responsible for inducing transition between the system states with energy being  shared or supplied by all three baths simultaneously.  Here $\sigma_z = |1\ra \la 1| - |0 \ra \la 0|$, $\sigma_x = |1\ra \la 0| + |0 \ra \la 1|$ are the standard Pauli spin operators with $|0\ra$ ($|1\ra$) representing the ground (excited) state of the qubit and $B_{\alpha}$ is the bath operator. $H_{B, \alpha} = \sum_{j} \omega_{\alpha, j} b_{\alpha, j}^{\dagger} b_{\alpha, j}$ is the standard bosonic bath Hamiltonian. The individual baths are in equilibrium with a fixed temperature. Note that, the Hamiltonian appearing in Eq.~(\ref{H-strong}) can be rigorously derived by performing a polaron transformation to the standard spin-boson Hamiltonian, bilinearly coupled to bosonic baths \cite{QAR-agar1}. 
Much like before, one can receive the mean currents and associated fluctuations following a full-counting statistics approach.  We follow Ref.~(\cite{QAR-agar1}) and present our results. 
The counting field dressed rate matrix can be written as,
\begin{equation}
\hat {\cal W}(\chi)=
\begin{pmatrix}
&-k_{0\to 1} & k_{1\to 0}^{\chi}  \\
& k_{0\to 1}^{\chi} & -k_{1 \to 0} &
\end{pmatrix}
\end{equation}
with the dressed transition rates given as,
\bea
k_{1\to 0}^{\chi} &=& \int_{-\infty}^{\infty} \frac{d\omega}{2\pi} \int_{-\infty}^{\infty}  \frac{d\omega'}{2\pi} M_h(\omega) M_c(\omega') M_w(\omega_0 \!-\! \omega\!-\!\omega') \nonumber \\
&&e^{-i \chi_h \omega} \, e^{-i \chi_c \omega'}\,e^{-i \chi_w (\omega_0-\omega-\omega')}
\label{rate-multi}
\eea
and $k_{0\to 1}^{\chi}= k_{1\to 0}^{\chi} (\omega_0 \to -\omega_0)$. The other rates in $\hat {\cal W}(\chi)$ correspond to $\chi=0$ limit. Here $M_{\alpha}(t) = \langle B_{\alpha}(t) B_{\alpha}(0) \rangle$ is the two-time bath correlation function with $B_{\alpha}(t) = e^{i H_{B,\alpha} t} \, B_{\alpha} \, e^{-i H_{B,\alpha} t}$ being the bath operator in the interaction picture.  $M_{\alpha}(\omega) = \int_{-\infty}^{\infty} ds \, M_{\alpha}(s) \, e^{i w s}$ is the corresponding Fourier transformed rates satisfying the detailed balance relation $M_{\alpha}(-\omega)= e^{-\beta_{\alpha} \omega} \, M_{\alpha}(\omega)$. Note the multiplicative nature of the individual rates appearing in Eq.~(\ref{rate-multi}). As shown in Ref.~(\cite{QAR-agar1}), various possible choices of $M_{\alpha}(\omega)$ can give rise to refrigeration and the optimal cooling efficiency $\ep_{\rm cool}$ is received in the tight-coupling limit which corresponds to highly engineered terminals with $M_{\alpha}(\omega)$ being described by a single frequency component.  Here we make the following choice for $M_{\alpha}(\omega)$ for $\omega>0$,
\bea
M_c(\omega) &=& \gamma_c \, \delta(\omega-\omega_c), \nonumber \\
M_h(\omega) &=& \frac{\gamma_h}{2\delta} \, \Big[ \Theta(\omega\!-\! \omega_h \!+\! \delta) \!-\!  \Theta(\omega\!-\! \omega_h \!-\! \delta)\Big], \nonumber \\
M_w(\omega) &=& \frac{\gamma_w}{2 \delta} \, \Big[ \Theta(\omega\!-\! \omega_w \!+\! \delta) \!-\!  \Theta(\omega\!-\! \omega_w \!-\! \delta)\Big].
\label{spectra-2level}
\eea
The negative frequency parts for the rates are fixed by the detailed balance equations. Here $\gamma_{\alpha} >0 $ is a dimensionless parameter, $\Theta(x)$ ($\delta(x)$) is the Heviside step (Dirac Delta) function. $\omega_{\alpha}$ is the central characteristic frequency for the terminal $\alpha$.  We set the central frequencies such that the resonant condition  $\omega_c + \omega_w = \omega_h$ is satisfied. The parameter $\delta$ (without the argument) appearing in $M_h$ and $M_w$ is the width parameter around central frequencies $\omega_h$ and $\omega_w$.  Note that $\delta \to 0$ corresponds to the tight-coupling or the optimal cooling limit here. In Fig.~(\ref{2-level-QAR}) we display our results for the bounds as a function of $\delta$. Expectedly, for small values of $\delta$ ($\delta \ll \omega_{\alpha}$) all the bounds saturate and exactly predict the values for both the mean cooling power and the corresponding mean cooling efficiency. For large $\delta$, i.e., beyond the tight-coupling limit, the hierarchy in the bounds for cooling power become transparent and once again within this hierarchy  ${\cal P}_w$ provides the tightest bound for the power. Moreover, interestingly, for these parameters, the additional universal bounds in Eq.~(9) for the cooling power and the cooling efficiency become tighter than the corresponding bounds obtained from the TURs.
{\it Summary.} We have shown that for autonomous absorption refrigerators operating in the linear response regime, a hierarchy in bounds exists for relative fluctuation of currents. We receive this result following the Onsager's reciprocity relation and additionally constraining the direction for the mean currents to realise a refrigerator. 
As a consequence, the independent looking universal bounds for cooling power that are predicated from the TURs, are not actually independent but also follow a hierarchy. Within this hierarchy, the best estimate for the cooling power is achieved when the bound is expressed in terms of work current fluctuation. We further provide completely independent universal bounds for cooling efficiency and cooling power that do not directly follow from the TURs and show for two paradigmatic ARs that these bounds can in fact become tighter and therefore can provide better estimates than ${\cal E}_{c}$ and ${\cal P}_w$. 


{\it Acknowledgements.} BKA acknowledges the financial support of the MATRICS grant MTR/2020/000472 from SERB, Government of India. BKA wants to thank the Shastri Indo-Canadian Institute for providing financial support for this research work in the form of a Shastri Institutional Collaborative Research Grant (SICRG). SM acknowledges support from the Council of Scientific \& Industrial Research (CSIR), India, (File number: 09/936(0273)/2019-EMR-I). SS acknowledges support from CSIR, India (Grant No. 1061651988).

\vspace{5mm}
\renewcommand{\theequation}{A\arabic{equation}}
\renewcommand{\thefigure}{A\arabic{figure}}
\renewcommand{\thesection}{A\arabic{section}}
\setcounter{equation}{0}  
\setcounter{figure}{0}

\appendix 

\section*{Appendix A: Bounds on cooling efficiency and power from the TUR}
In the main text, we provide bounds on  cooling efficiency and cooling power following the TUR relations. Here we provide the derivation. We write the net EP rate in the steady-state as $\la \sigma \ra=- \sum_{\alpha} \beta_{\alpha} \la j_{\alpha} \ra$ which can be expressed in terms of the currents involving the cold and work terminals by following the conservation of steady-state currents  i.e., $\sum_{\alpha} \la j_{\alpha} \ra =0$. One then receives,
\bea
\la \sigma \ra &=& \la j_c \ra \, \big(\beta_h \!-\! \beta_c) +  \la j_w \ra \, \big(\beta_h \!-\! \beta_w) \nonumber \\
&=& \big(\beta_c-\beta_h) \Big[ \frac{\ep_{\rm cool}}{\la \ep \ra} -1 \Big] \la j_c \ra \geq 0.
\eea
The positivity is ensured by the second law of thermodynamics.  Now, since $\la j_c \ra >0$ and $\beta_c >\beta_h$, we receive $\la \ep \ra \leq \ep_{\rm cool}$. Combining the above expression with the TUR for the cold current i.e.,
\be
\la \sigma \ra \frac{D_{c}}{\la j_{c} \ra^2} \geq 1,
\ee
we receive, 
\be
\frac{\ep_{\rm cool}}{\la \ep \ra} \geq 1 + \frac{\la j_c \ra}{(\beta_c-\beta_h) D_c}
\ee
which gives a lower bound on the cooling efficiency, 
\be
\langle \epsilon \rangle \leq \epsilon_{\rm cool} \Big{/} \Big(1 + \frac{\la j_c \ra }{(\beta_c\!-\!\beta_h) \, D_c}\Big) \leq \ep_{\rm cool}.
\ee
and matches with Eq.~(\ref{eff-bound-udo}).   
One can alternatively express the above bound on the cooling efficiency as a bound on the cooling power, i.e.,
\be 
 \la j_c \ra \leq (\beta_c \!-\! \beta_h) \frac{\Big(\ep_{\rm cool} \!-\! \la \ep \ra\Big)}{\la \ep \ra} \, D_c \equiv  {\cal P}_c
\ee
The other bounds on the cooling power can be similarly received following the TURs for hot and work currents. 

\renewcommand{\theequation}{B\arabic{equation}}
\setcounter{equation}{0}  
\section*{Appendix B: Equality for TUR in the tight-coupling limit}
In this section we show that for the multi-affinity AR setup, the TUR bound saturates in the tight-coupling limit. We perform the calculation, once again for the cold terminal which can be trivially extended for the other terminals also. In the linear response regime, following Eq.~(5), we write the TUR relation as \cite{Barato:2015:UncRel}, 
\be
\la \sigma \ra \frac{D_{c}}{\la j_{c} \ra^2} =1 + \frac{1}{\la j_c \ra^2} \sum_{\alpha, \gamma=c,h} \Big(L_{cc} L_{\alpha \gamma} - L_{c \alpha} L_{c \gamma}\Big)  \,  A_{\alpha}^{w} \,  A_{\gamma}^{w},
\label{TUR-linear}
\ee
where  $A_{c}^{w}= (\beta_w \!-\! \beta_c)$ and $A_{h}^{w}= (\beta_w \!-\! \beta_h)$ and the EP rate is 
\be
\la \sigma \ra = \sum_{\alpha, \gamma=c, h} L_{\alpha \gamma} \,  A_{\alpha}^{w} \,  A_{\gamma}^{w}.
\ee
It is easy to check that the only non-zero term in the above summation in Eq.~(\ref{TUR-linear}) corresponds to $\alpha=\gamma=h$, then one can rewrite above equation as,
\be
\la \sigma \ra \frac{D_{c}}{\la j_{c} \ra^2} =1 + \frac{1}{\la j_c \ra^2} \Big(L_{cc} L_{hh} - L^2_{c h}\Big)  \,  \big(A_{h}^{w}\big)^2  \geq 1
\label{TUR-linear-1}
\ee
where the equality is reached in the tight-coupling regime i.e., when the determinant of the Onsager's matrix vanishes. This result also follow from the two other TURs.

\renewcommand{\theequation}{C\arabic{equation}}
\setcounter{equation}{0}  

\section*{Appendix C: Proof for $\sqrt{\frac{D_c}{D_w}} \leq \ep_{\rm cool}$}
In the linear response regime, writing the currents $\la j_c \ra$ and $\la j_w \ra$ in terms of the Onsager's coefficients $\bar{L}^{h}_{\alpha \gamma}, \alpha,\gamma=w,c$ and imposing the refrigeration conditions i.e., $\la j_w\ra A_w^{h} \geq 0$ and $\la j_c\ra A_c^{h} \leq 0$, we receive the following inequality
\be
\bar{L}_{cc} \, \big(A_c^{h} \big)^2 \leq \bar{L}_{ww} \, \big(A_w^{h} \big)^2.
\ee
where $A_{\alpha}^{h} = (\beta_h -\beta_{\alpha})$. 
As a result, we immediately receive,
\be
\langle \ep \rangle \leq \sqrt{\frac{D_c}{D_w}} \equiv \sqrt{\frac{L_{cc}}{L_{ww}}} \leq \frac{(\beta_h\!-\!\beta_w)}{(\beta_c\!-\!\beta_h)} = \ep_{\rm cool}
\ee

\renewcommand{\theequation}{D\arabic{equation}}
\setcounter{equation}{0}  
\setcounter{figure}{0}

\section*{Appendix D: {Quantum Absorption Refrigerators with a three level working fluid under weak-coupling}}
\begin{figure}
\includegraphics[trim=50 300 50 300, clip, width=0.8\columnwidth]{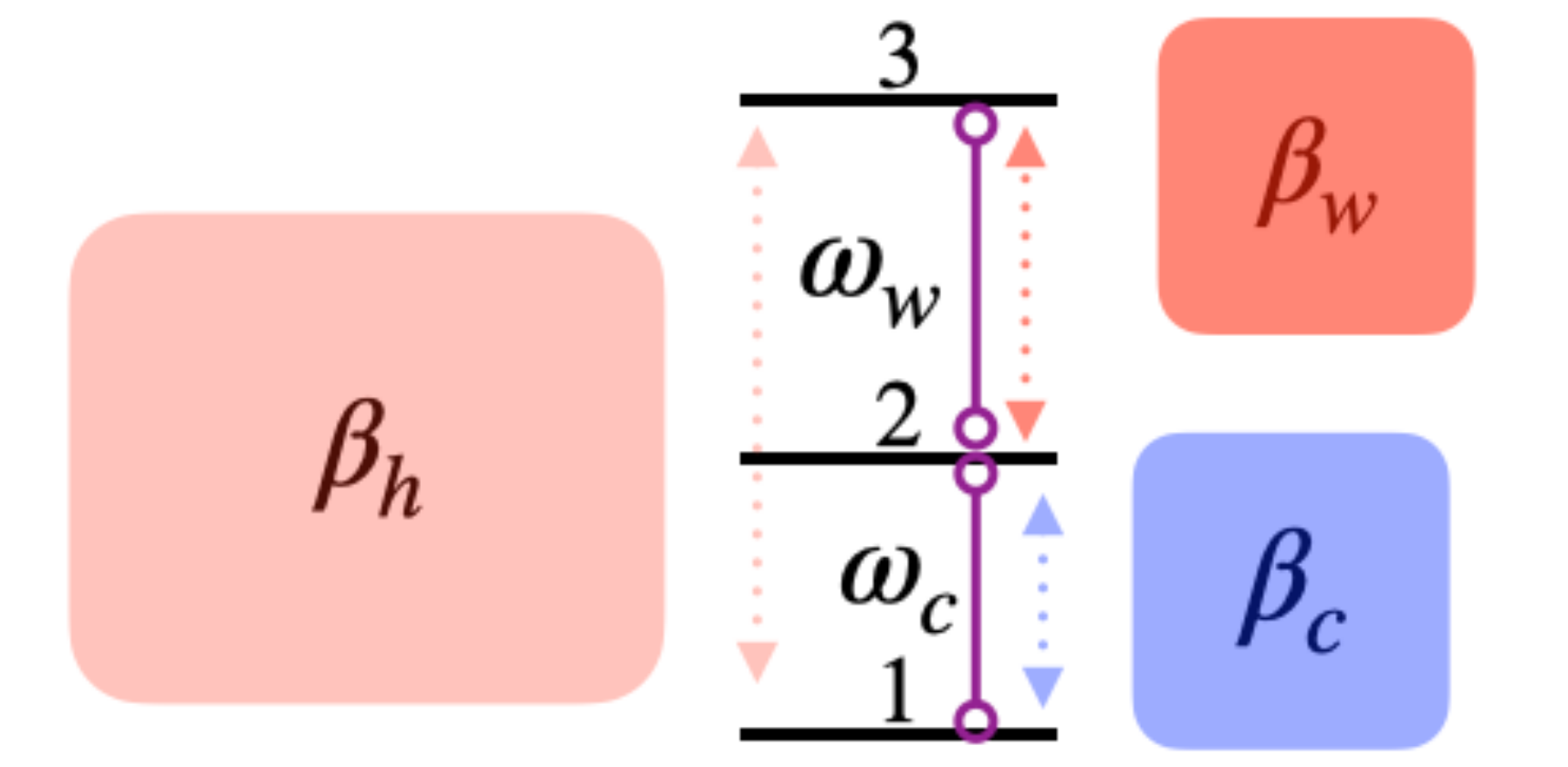}
\caption{(Color online) : Schematic of a minimal quantum AR  in the weak-coupling regime where a three-level working fluid is coupled weakly to  three different terminals. The baths induced incoherent transitions are shown by the dashed arrows. The energy gaps between the states are indicated by solid lines. }
\label{3-level-QAR}
\end{figure}

In the main text, we presented results for a four-level quantum AR in the tight-coupling limit ($g \ll \omega_c)$.
In this section, we choose a relatively simple three level system, shown in Fig.~(\ref{3-level-QAR}), and show that, while working as a quantum AR in the weak-coupling regime,
the relative fluctuation for all the three currents are the same and occurs due to the tight-coupling situation. Recall that, this limit corresponds to the equality in the bounds presented in the main text.  Note that, in the tight-coupling limit ($g \ll \omega_c$), the cooling efficiency for four-level quantum AR matches with the three-level weakly coupled quantum AR. However, the cooling power can be different.  Interestingly, for this three level setup, our proof for the equality in relative fluctuation does require the explicit forms for the incoherent rates induced by the baths. The central ingredient here is the additivity feature of the rates. 

Similar to the four-level setup, the counting-field dressed rate matrix that describes the population dynamics for the system states can be written down using the standard quantum master equation approach \cite{Segal-abs-2}. This is given as,
\bea \hat {\cal W}(\chi)=
\begin{pmatrix}
-k_{1\rightarrow 2}^c-k_{1\rightarrow 3}^h  &  k_{2\rightarrow 1}^c e^{-i\chi_c\omega_c}   & k_{3\rightarrow 1}^h  e^{-i\chi_h\omega_h}  \\
k_{1\rightarrow 2}^c e^{i\chi_c \omega_c}  &   -k_{2\rightarrow 1}^c - k_{2\rightarrow 3  }^w & k_{3\to 2}^w e^{-i\chi_w\omega_w}\\
k_{1\rightarrow 3}^h  \,  e^{i\chi_h\omega_h}  &   k_{2\rightarrow 3}^w e^{i\chi_w\omega_w} &  -k_{3\rightarrow 2 }^w -k_{3\to 1}^h\\
\end{pmatrix}.
\nonumber\\
\label{eq:M}
\eea
%
%
%
As shown in the figure, the energy states are labeled from bottom to top by 1,2,3, with energy gaps denoted by $\omega_c$ and $\omega_w$. 
As mentioned earlier, we do not need to specify the explicit form of these transition rates. 
The statistics for all three currents is determined by the largest eigenvalue (real part) of this characteristic polynomial. The characteristic polynomial for the largest eigenvalue satisfy the equation 
\be
\lambda_{\rm max}^3 - a_1 \, \lambda_{\rm max}^2 + a_2 \, \lambda_{\rm max} - a_3 (\chi)=0
\label{charc}
\ee
with the condition that $\lambda_{\rm max}(\chi=0) = 0$.  Here, 
\bea
a_1 &=& w_{0,0} + w_{1,1} + w_{2,2}, \nonumber \\
a_2 &=& w_{0,0} w_{1,1} + w_{0,0} w_{2,2} + w_{1,1} w_{2,2} - w_{0,1}(\chi_c)\,w_{1,0}(\chi_c)\nonumber \\
&& - w_{0,2}(\chi_h)\,w_{2,0}(\chi_h) - w_{1,2}(\chi_w)\,w_{2,1}(\chi_w), \nonumber \\
a_3(\chi) &=& w_{0,0} \, w_{1,1}\, w_{2,2} - w_{0,0}\, w_{1,2}(\chi_w) \, w_{2,1}(\chi_w) - w_{0,1}(\chi_c) \times \nonumber \\
&&w_{1,0}(\chi_c) \,w_{2,2} +  w_{0,1}(\chi_c)\,  w_{1,2}(\chi_w)\, w_{2,0} (\chi_h) \nonumber \\
&+&  w_{0,2} (\chi_h)  \, w_{2,1}(\chi_w) \, w_{1,0}(\chi_c) - w_{0,2} (\chi_h) \, w_{2,0}(\chi_h) \,w_{1,1}, \nonumber \\
\eea
where $w_{i,j}$ are the elements of the matrix $\hat {\cal W}(\chi)$ \cite{Segal-abs-2}. Notice that, $a_1$ is counting field independent. Moreover, the counting field dependent phase factors exactly cancels out in $a_2$. As a result, $a_3$ is the only counting field dependent term in Eq.~(\ref{charc}). 
We can obtain analytical expression for the current for bath $\alpha$ as 
\bea
\la J_{\alpha} \ra &=& \frac{1}{a_2} \, \frac{\partial{a_3}}{\partial(i x_{\alpha})} \Big {|}_{\chi=0}, \quad \alpha=h,c,w.
\eea
It can be easily checked that the currents are proportional to each other and given by  
\bea
\la j_c \ra &=& \frac{\omega_c}{a_2} \Big[ w_{0.2} \, w_{2,1} \, w_{1,0} - w_{0,1} \, w_{1,2} \, w_{2,0} \Big], \nonumber \\
\la j_w \ra &=& \frac{\omega_w}{\omega_c} \, \la j_c \ra, \nonumber \\ 
\la j_h \ra &=& -\frac{\omega_c+\omega_w}{\omega_c} \, \la j_c \ra.
\eea
The corresponding noise for bath $\alpha$ is given as
\be
D_{\alpha}  = \frac{\la \la J_{\alpha}^2 \ra \ra}{2} = \frac{1}{2\, a_2} \, \Big[ \frac{\partial^2{a_3}}{\partial(i x_{\alpha})^2}  + \frac{2 a_1}{a_2^2} \Big(\frac{\partial{a_3}}{\partial(i x_{\alpha})}\Big)^2\Big] \Big{|}_{\chi=0}. \,\,\, 
\ee
As a result, the relative fluctuation simplifies to
\bea
\frac{D_{\alpha}}{\la J_{\alpha}\ra^2} = \frac{a_1}{a_2} + \frac{ a_2}{2} \, \frac{ a_3^{''}}{(a_3^{'})^2},
\eea
where the primes indicate the order of the derivative with respect to the respective counting field  $\chi_{\alpha}$. Notice that $a_1$ and $a_2$ remains the same  for any current. Interestingly, it turns out the second term $a_3^{''}/(a_3^{'})^2$ is also the same for any current and is given by
\be
\frac{ a_3^{''}}{(a_3^{'})^2} = \frac{ w_{0,2}\, w_{2,1}\, w_{1,0} \!+\! w_{0,1}\, w_{1,2}\, w_{2,0}  } {\big( w_{0,2}\, w_{2,1}\, w_{1,0} \!-\! w_{0,1}\, w_{1,2}\, w_{2,0} \big)^2}.
\ee
We therefore see that the relative fluctuation of currents for all three terminals are the same, i.e.,
\be
 \frac{D_c}{\la j_c \ra^2} = \frac{D_h}{\la j_h \ra^2} = \frac{D_w}{\la j_w \ra^2}.
\ee
This completes our proof.   

\begin{figure}[t]
\includegraphics[trim=70 240 100 240, clip, width=0.8\columnwidth]{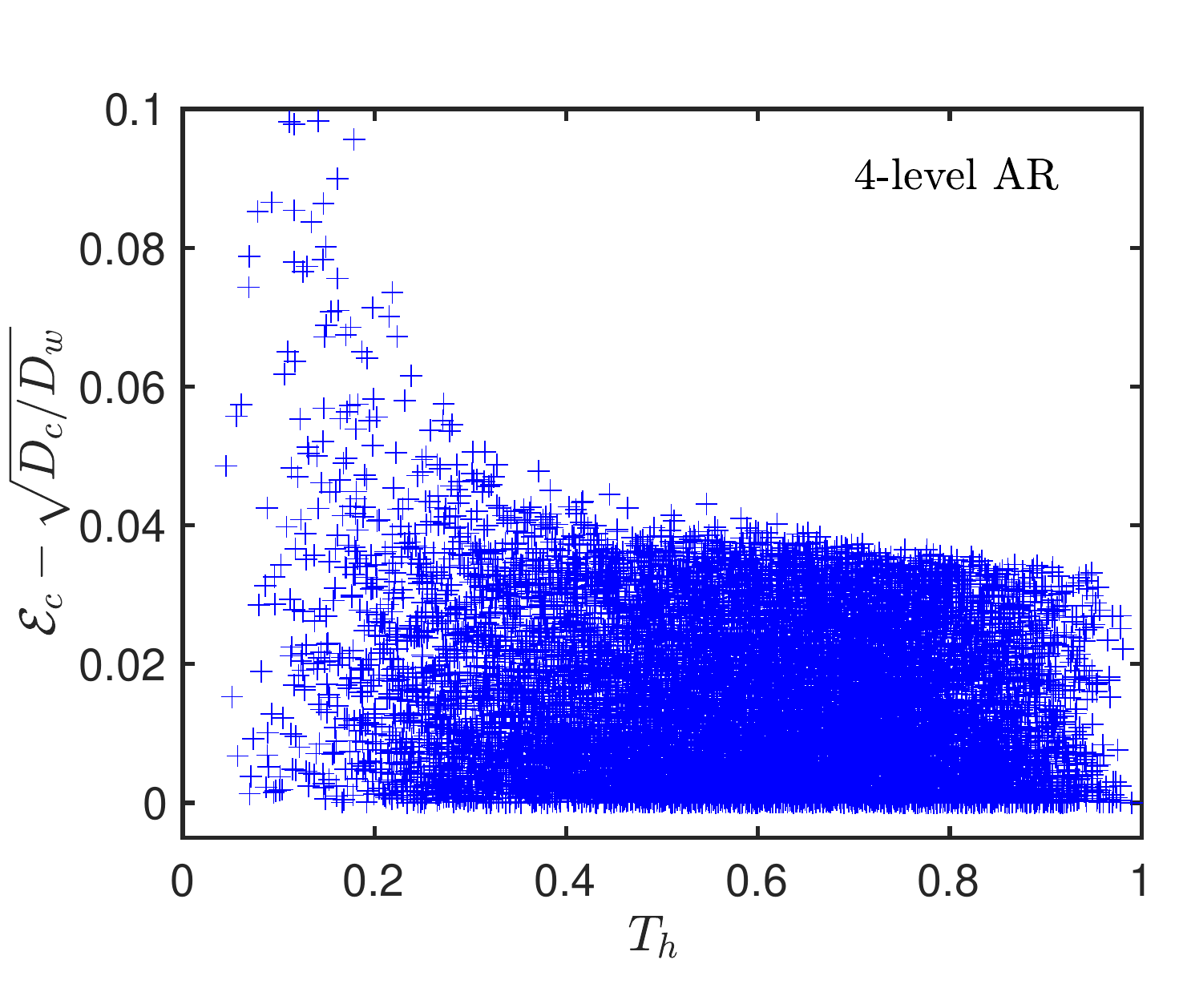}
\caption{(Color online) : Scatter plot of the difference in the bounds ${\cal E}_c - \sqrt{D_c/D_w}$ for the 4-level quantum AR. We simulate the system over a broad parameter regime and observe that ${\cal E}_c - \sqrt{D_c/D_w}>0$ which implies that the bound $\la \ep \ra \leq  \sqrt{D_c/D_w}$ can be tighter compared to ${\cal E}_c$. We perform the simulation over 1 million random points and select those data points for which we realise a refrigerator. Here we choose the temperature of all three baths within the interval $[0.1,1]$ randomly and set the linear response regime by demanding that the ratio $\Delta T/T $  is always smaller than $0.05$ where $\Delta T$ is the temperature difference between any two terminals and $T$ is the average temperature for the same two terminals. The value of $g$ is chosen within the interval $[0,0.05]$.  We kept other parameters fixed  with values for $\omega_c = 0.3$ and $\omega_w=0.6$.}
\label{4-level-QAR}
\end{figure}

\begin{figure}[t]
\includegraphics[trim=70 240 100 240, clip, width=0.8\columnwidth]{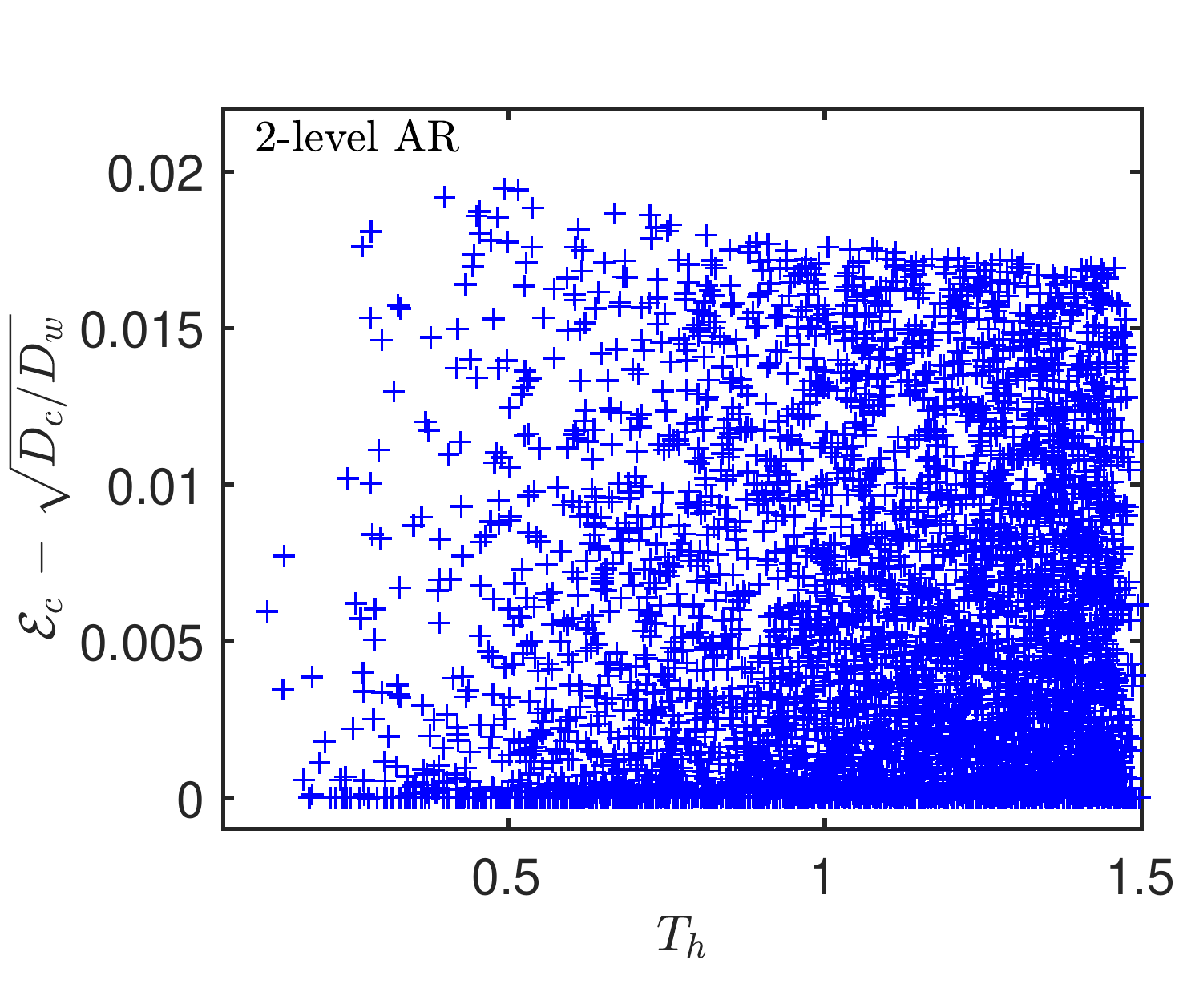}
\caption{(Color online) : Scatter plot of the difference in the bounds ${\cal E}_c - \sqrt{D_c/D_w}$ for the 2-level quantum AR. We perform numerical simulation over a broad parameter regime by choosing the different model parameters randomly. We perform the simulation for 10 million points. Here we choose the temperature of all three baths within the range $[0.1,1.5]$ randomly. The value of $\delta$ is chosen within the interval $[0,0.5]$. Once again, we fix the linear response regime by demanding that the ratio $\Delta T/T $  is always smaller than $0.05$ where $\Delta T$ is the temperature difference between any two terminals and $T$ is the average temperature for the same two terminals. We observe that in the linear response regime ${\cal E}_c - \sqrt{D_c/D_w}>0$ implying  $\la \ep \ra \leq  \sqrt{D_c/D_w}$ can become tighter compared to ${\cal E}_c$. }
\label{2-level-QAR}
\end{figure}

\renewcommand{\theequation}{E\arabic{equation}}
\setcounter{equation}{0}  
\section*{Appendix E: {Test for bounds ${\cal E}_c$ obtained from TUR and the bound $\sqrt{D_c/D_w}$ in Eq. (9)}}
In this section, we perform a comparison study between the bounds for cooling efficiency ${\cal E}_c$ in Eq.~(2), as received from the TUR, and the bound that followed from the hierarchy of the relative fluctuations between the cold and the work terminal i.e., $\sqrt{D_c/D_w}$ , given in Eq.~ (9). For the 4-level and 2-level quantum AR's we provide scatter plots in Fig.~(\ref{4-level-QAR}) and Fig.~(\ref{2-level-QAR}) respectively, by choosing different model parameters randomly. For both these models, in the linear response regime, we observe the bound $\sqrt{D_c/D_w}$ is always tighter than the bound ${\cal E}_c$ i.e., the difference ${\cal E}_c-\sqrt{D_c/D_w}$ is always positive.

\end{document}